\documentclass[aps,showpacs,prl]{revtex4}
\usepackage{graphicx,float}

\begin{document}

\preprint{}


\count255=\time\divide\count255 by 60
\xdef\hourmin{\number\count255}
  \multiply\count255 by-60\advance\count255 by\time
 \xdef\hourmin{\hourmin:\ifnum\count255<10 0\fi\the\count255}

\newcommand{\xbf}[1]{\mbox{\boldmath $ #1 $}}

\title{New near-threshold  mesons}

\author{Thomas D. Cohen}
\affiliation{Department of Physics, University of Maryland,
College Park, MD 20742-4111, USA}
\author{Boris A. Gelman}
\affiliation{Department of Physics,
 University of Arizona, Tucson, AZ 85721, USA}

\author{Shmuel Nussinov}
\affiliation{School of Physics and Astronomy, 
Tel Sackler Faculty of Exact Sciences, Tel Aviv University,
Tel Aviv, Israel \\ and \\ 
Department of Physics and Astronomy, University of South Carolina,
SC 29208, USA}


\date{September 2003}

\def\D{D^{*}_{sJ} (2317)}
\def\r{\rangle}
\def\th{\Theta(+)}
\def\aa{a_{0}(980)}
\def\ff{f_{0}(980)}
\def\a{a_{0}}
\def\f{f_{0}}
\def\e{\epsilon}
\def\P{{\cal P}_i}
\def\Ha{{\cal H}_a}
\def\Hb{{\cal H}_b}

\begin{abstract}

We show that under a number of rather plausible assumptions QCD spectrum 
may contain a number of mesons which have not been predicted or
observed.  Such states will have the quantum numbers of two
existing mesons and masses very close to the
dissociation threshold into the two mesons. Moreover, at least
one of the two mesonic constituents itself must be very close to its
dissociation threshold. In particular, one might expect the existence of 
loosely bound systems of $D$ and $\D$; similarly, $K$ and $\ff$, $\bar{K}$ and 
$\ff$, $K$ and $\aa$ and $\bar{K}$ and $\aa$ can be bound. 
The mechanism for binding in  these
cases is the S-wave kaon exchange.  The nearness of one of the
constituents to its decay threshold into a kaon plus a remainder,
implies that the range of the kaon exchange force becomes
abnormally long---significantly longer than $1/m_K$ which
greatly aids the binding.

\end{abstract}

\pacs{12.38.Aw, 12.40.Yx, 14.80.-j, 21.30.Fe, 21.45.+v}



\maketitle

\section{Introduction}

The particle data book abounds with hadronic resonances \cite{PDG}. However,
there are comparatively  few states which are very close to the
threshold for decay into other mesons.  Recently a new near-threshold 
state---the narrow $\D$ ($I=0$ and possibly $J^P=0^+$) at about $40\, MeV$
below $KD$ threshold---was found at BABAR \cite{BB}, CLEO \cite{CL} and
Belle \cite{BL}.

Note, that the $\D$ state can be interpreted in a number of ways: (a) as the 
missing triplet S-wave
$(J^P=0^+)$ $|c\bar{s} \rangle$ ``quarkonium state''; (b) as a
``single bag'' $\left (|c\,\bar{s}\,u\,\bar{u}\rangle +
|c\,\bar{s}\,d\,\bar{d}\rangle \right ) /\sqrt{2}$ isosinglet
state; or (c) as an isosinglet ``molecular'' bound state $\left
(|K^+\, D^0\r + |K^0\, D^+ \r \right )/\sqrt{2}$ of two separate
hadrons.   The two hadrons in the last case can be bound---just
like the deuteron---by an attractive potential due to the
t-channel exchange of various light mesons. The Lagrangian has
``off-diagonal'' terms such as $q\,\bar{q}$ pair creation and
annihilation and/or ``bag'' fissioning and rejoining
interconnecting states of type (a) and (b), and (b) and (c)
respectively. As a result we expect that $\D$ is a superposition
of all three states in (a), (b) and (c). The question is then
which one dominates the state $|\D \r$.

Regardless of how one chooses to interpret the state there is one
key fact about this state which will play a major role in what
follows: the state is extremely close to the $KD$ threshold.
This situation parallels a case of the pseudoscalar isosinglet
and isotriplet mesons---$\ff$ and $\aa$---which are very close to
the  $K\bar{K}$ threshold. These states can correspond to any
one of the three cases above  provided the quark pair $c\bar{s}$ is
replaced by $s\bar{s}$ \cite{RJ}.

In Ref.~\cite{SN} one of us argued in favor of interpretations (b)
and (c). The argument in \cite{SN} was based on the fact that the
mass difference of approximately $20 \, MeV$ between $\D$ and the
state $D^{*}_{0}$ $(J^P=0^+)$ with a mass of about $2300\, MeV$
(BELLE \cite{Dq}) is significantly smaller than an approximate
$100\, MeV$ split between any two ``strangeness analogue''
$X_{\bar{s}}-X_{\bar{q}} \ (q=u, \, d$) mesonic or baryonic state
\cite{NS}. Likewise, the isotriplet ($P$-wave)  $s\bar{s}$
state is $40\, MeV$ lighter than the isotriplet $S$-wave $\phi
(1020)$ rather than being more than $350 \, MeV$ heavier, as is
the case for all other nonets. This is an argument against the
``minimal'' interpretation of $\ff$ and $\aa$ states as $s\bar{s}$ pairs.
To the extent that $\ff$, $\aa$ and $\D$ are indeed of
type (b) or (c) then the following prediction can be made.  A ``QCD
inequality'' \cite{NL} implies yet another pseudoscalar $c
\bar{c}$ state approximately $100\, MeV$ below the threshold
\cite{SN}. This state can be discovered via the $\eta \eta_c$
decay mode in BABAR and Belle. Ordinary $c \bar{c}$ states are
accounted for and such a state would have to be interpreted as
being exotic.

Of course, one can take a far more agnostic position as far as
the interpretation of the $\D$ or the $\ff$ and $\aa$.  Since
the three interpretations were expressed in terms of model
concepts rather than QCD degrees of freedom, one can argue that
even in principle there is no way to distinguish between them.
However one chooses to interpret these states,  we can rely on the
fact that they have $J^P=0^+$ and are only very slightly below the
corresponding break-up thresholds: $40 - 50 \, MeV$ below $KD$ and
$10 - 20 \, MeV$ below $K\bar{K}$  thresholds respectively.  This fact
greatly facilitates
the possibility that these mesons will be bound weakly
into ``molecular''-like states: $|D \D \r$ and  $|K \f\r$, $|K\a \r$, 
$|\bar{K}\f \r$, $|\bar{K}\a \r$.  While any of these
states would be interesting to observe, the $|\D \, D\r$ is of
particular interest owing to the fact that by quantum numbers
alone ($S=1, \, C=2$) it is manifestly exotic.

\begin{figure}[ht]
\begin{minipage}[t]{0.31\columnwidth}
\centering
\includegraphics[height=5cm]{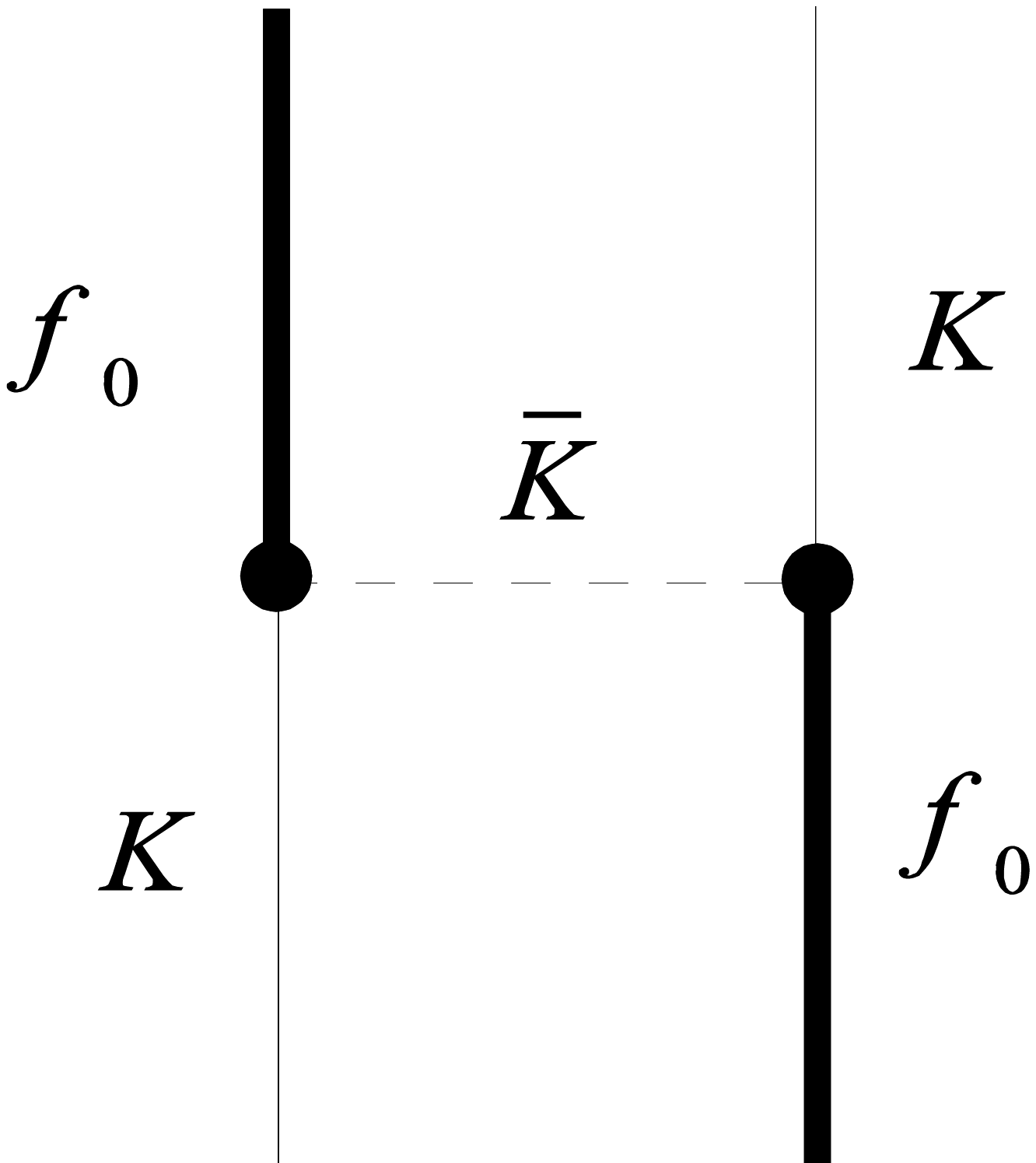}
\caption{$|K \f \r \ \left (|\bar{K} \f \r \right )$
bound state:\\ $t$-channel $K$ exchange.}
\label{fig1}
\end{minipage}
\hfill
\begin{minipage}[t]{0.31\columnwidth}
\centering
\includegraphics[height=5cm]{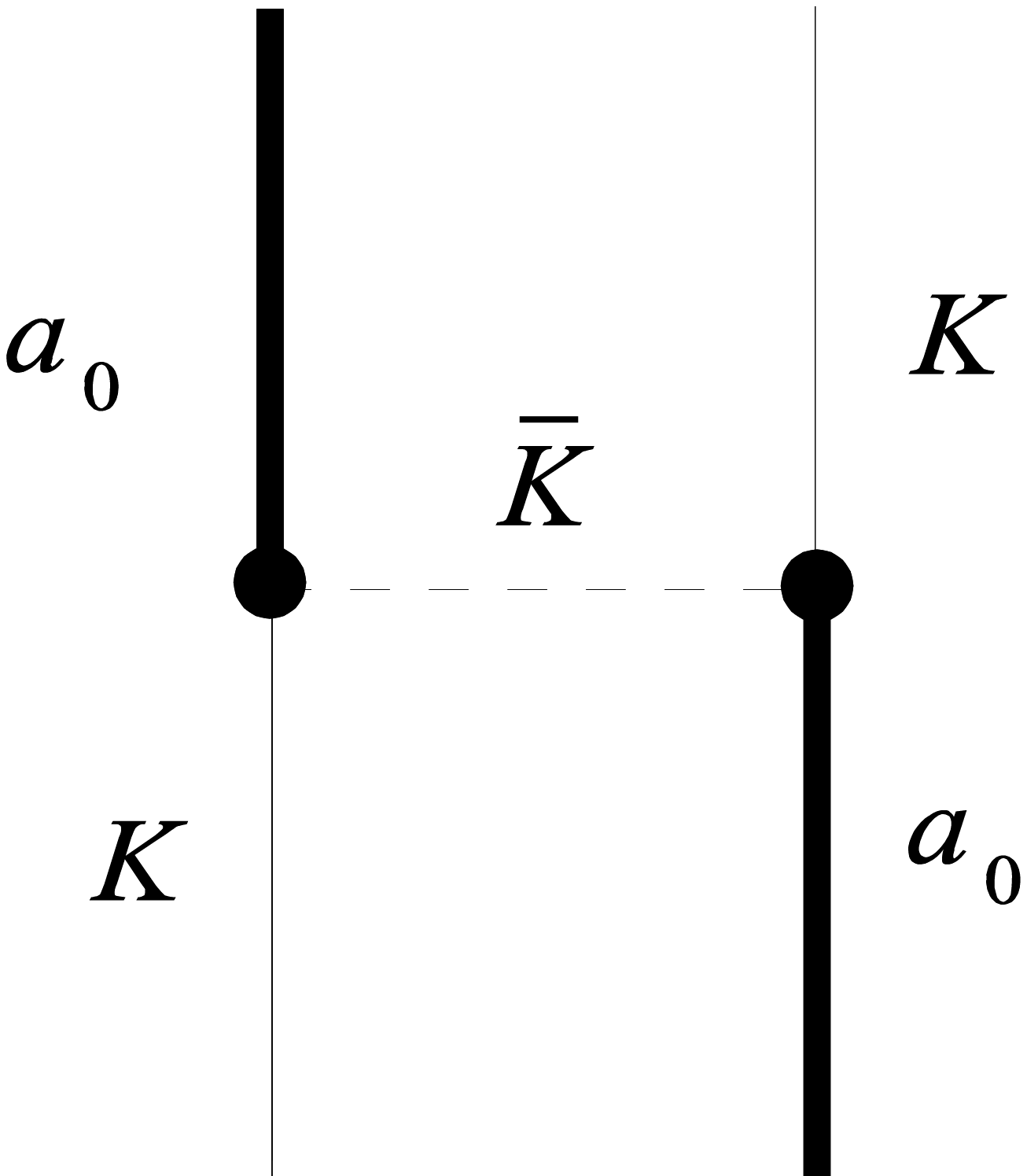}
\caption{$|K \a \r \left ( |\bar{K} \a \r \right )$
bound state:\\ $t$-channel $K$ exchange.}
\label{fig2}
\end{minipage}
\hfill
\begin{minipage}[t]{0.31\columnwidth}
\centering
\includegraphics[height=5cm]{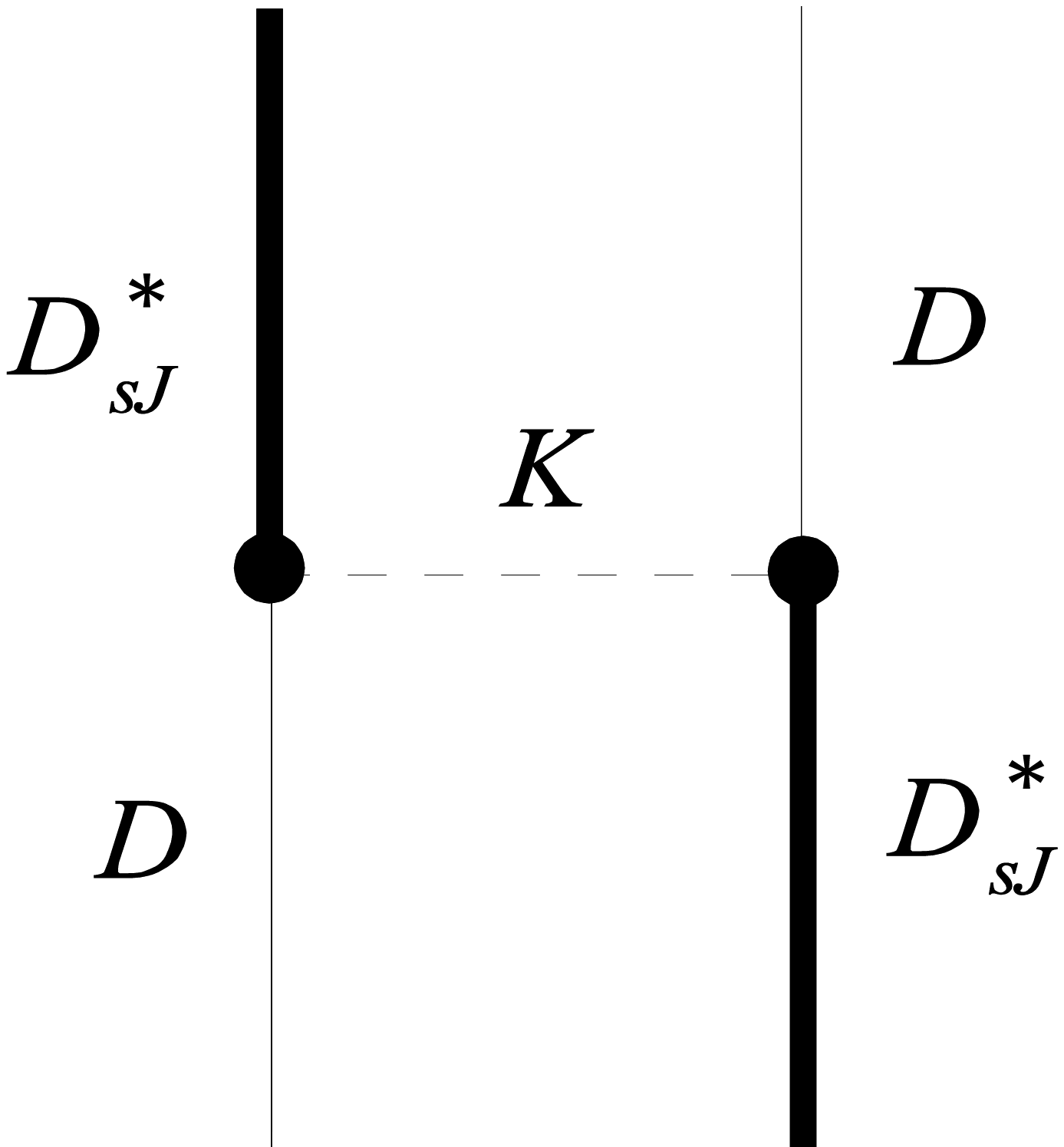}
\caption{$|D \D \r$ bound state: $t$-channel $K$ exchange.}
\label{fig3}
\end{minipage}
\hfill
\end{figure}

\section{The binding mechanism}

The key to our analysis is the following fact.  If there exists a
meson $\Ha$ quite near to the break-up threshold into a $K$ plus
another meson, $\Hb$, then there will be an unnaturally
long-ranged force between mesons $\Ha$ and $\Hb$ due to $K$
exchange.  Now suppose the $K$-exchange force leads to an
attractive $S$-wave interaction.  This in turn will lead to tendency
toward binding.  Of course, if the state formed is deeply bound,
then the fact the interaction is comparatively long ranged plays
no important role.  Indeed, if the state is deeply bound, the
$K$-exchange mechanism clearly does not dominate and one would
need to understand the physics at the quark level. However, if the
state formed is relatively loosely bound, the unnaturally
long-ranged character of the potential becomes essential  to the
binding. As we show below, for a large range of parameters, this
is precisely what occurs: the $K$-exchange is strong enough to
bind the two mesons together but is weak enough so that the
resulting bound state is dominated by ranges where the
$K$-exchange is expected to be important.  Note, that if we are in
such a weakly bound region, it is legitimate to use a
nonrelativistic Schr\"odinger equation to describe the dynamics.

We begin the analysis by deriving the interaction due to the $K$-exchange. 
Consider two ``heavy'' mesons ${\cal H}_a$ and ${\cal
H}_b$ ($\Ha = \f$ ($\a$)--case (1)---or $\D$---case (2); $\Hb =
K$ ($\bar{K}$)---case (1)---or $D$---case (2))  of the same spin and opposite
parity. The mesons
need only be heavy in the sense that their masses are much larger
than the ultimate binding energy between them.  In this regime,
the binding potential, $V(r)$ ($r=|r_a-r_b|$), can be obtained from a
one-kaon exchange amplitude shown in the Feynman diagrams of
Figures~\ref{fig1}, \ref{fig2}, \ref{fig3}.  They can be evaluated in the
limit in which
the recoil energies of the mesons ${\cal H}_a$ and ${\cal H}_b$
are neglected (such a limit is justified near the ${\cal H}_a \,
{\cal H}_b$ threshold). In this limit the energy transfer carried by an
off-shell $K$ meson is $\epsilon =
m_b+m_K-m_a$, where $m_a,\, m_b ,\, m_K$ are the masses of
${\cal H}_a$, ${\cal H}_b$ and $K$  mesons respectively.
In the cases of interest here this
energy is $\epsilon \lesssim 50 \, MeV$, {\it i.e.} much smaller than
the mass of $K$ meson. The $\epsilon$ can be viewed
as the binding energy of $K$ and $\Hb$ into $\Ha$.

Keeping the leading term in $1/m_K$ expansion of the
kaon propagator and taking the Fourier transform of the amplitude
one obtains an attractive Yukawa-like potential:
\begin{equation}
V(r)=-\,{{\rm g}^{2}_{i} \over 16 \pi \, m_a\, m_b} \,\,
{{\rm exp}\left [-r\, \sqrt{2m_K \ \e_i} \right]\over r} =
\alpha_i \,\, {{\rm exp}\left [-\kappa_i \, r\right]\over r}\, ,
\label{V}
\end{equation}
where ${\rm g}_i$
(${\rm g}_1 = {\rm g}_{K\bar{K}\f(\a)}$,
${\rm g}_2 = {\rm g}_{KD\D}$)
is (the mass dimension two) coupling constant of the $S$-wave
Yukawa coupling $K\Ha\Hb$. The factor
$4 \ m_a \ m_b$ in the denominator in Eq.~(\ref{V}) comes from the
non-relativistic normalization of the scalar wave functions of
${\cal H}_a$ and ${\cal H}_b$. Consequently, the coupling
constant $\alpha_i =-\,{\rm g}^{2}_{i} /\left ( 16 \pi \, m_a\, m_b \right)$
is dimensionless.

The potential in Eq.~(\ref{V}) can be interpreted as the
(asymptomatic) profile function of the field strength of the
virtual $K$ inside the $H_a$ bound state (up to the coupling
constants). It has the form of an outgoing spherical wave with a
purely imaginary momentum $k_i=i\, \kappa_i$ with $\kappa_i$
equal to $\sqrt{2\, m_K \, \e_i}$. Note that the $\kappa_i$ in
Eq.~(\ref{V}) replaces $m_K$ in the standard Yukawa-like potential
yielding much longer range potentials. This effect is huge for
possible $DD^{*}$  loosely state considered by
T$\ddot{\rm{o}}$rnqvist \cite{TO}. As in our case, $\e=m_D
+m_{\pi}-m_{D^*}$ is tiny. However, in that case, the interaction is
in $P$-wave with a derivative $\pi D D^{*}$ coupling. As a result,
the increase in
the range in this case is essentially compensated by the corresponding
decrease in the strength of the coupling. This is not the case
for the $S$-wave momentum independent couplings relevant in our
case.

\section{Estimated binding energies}

The central result of this paper is that for a wide range of
``reasonable'' interactions between $\Ha$ and $\Hb$ binding results.

In the two cases considered here---(1) $|K\f \r$, $|K\a \r$, $|\bar{K}\f \r$,
$|\bar{K}\a \r$ and
(2) $|D\D \r$---the values of $\kappa_i$ are:
\begin{equation}
100 \, \lesssim  \kappa_1 \lesssim 140 \,\, MeV \,, \,\,\,\,\,\,\,\,\,\,
200 \,  \lesssim  \kappa_2 \lesssim 220 \,\, MeV \,.
\label{kappa12}
\end{equation}
The variation in Eq.~(\ref{kappa12}) is due to the
differences in binding energies for various $D$ and $K$ charge sates.

The binding energies of the $|\Ha \Hb \r$ ``molecules'' can now be determined
from the Schr\"odinger
equation with a potential given in Eq.~(\ref{V}) and reduced masses
$\mu_1  =  2 \ m_K/3 \approx 330 \, MeV$ (case (1)) and
$\mu_2 \approx 1030 \, MeV$ (case (2)).
The binding energies and the typical sizes of the ground state wave functions
(given by $\sqrt{<r^2>}$) for a number of couplings $\alpha_i$ and values
of $\kappa_i$ (Eq.~(\ref{kappa12})) are shown in
Tables~\ref{tabl1}, \ref{tabl2},\ref{tabl3} (case (1)) and
Tables~\ref{tabl4}, \ref{tabl5},\ref{tabl6} (case (2)) \cite{LS}.

In Ref.~\cite{CI} the value of the coupling constant
${\rm g}^{2}_{K\bar{K}\f}/(4\pi)$ was determined to be $0.6 \ GeV^2$.
The corresponding dimensionless coupling is
$\alpha_{K\bar{K}\f}=-\ {\rm g}^{2}_{K\bar{K}\f}/(16\pi \ m^{2}_{K}) \approx
-\ 0.6$. We also assume that the same value is
applicable in the case of $K\a$ and $D\D $ systems.

\begin{table}[ht]
\begin{minipage}[t]{0.48\columnwidth}
\begin{tabular}{|c|c|c|}
\hline
$\alpha$ & ${\rm E_{B},}\,\, MeV$ & $\sqrt{<r^2>},\,\, fm$ \\
\hline       $-0.4$ & $-2.9 $         & $36.1$   \\
\hline       $-0.6$ & $-17.3$         & $17.1$   \\
\hline       $-0.8$ & $-44.4$         & $11.8$   \\
\hline
\end{tabular}
\caption{$|K \ \f \r$ ($|K \ \a \r$): the binding energies and
the size of the ground sate wave function for the potential in
Eq.~(\ref{V}) with $\kappa=100 \ MeV$.} \label{tabl1}
\end{minipage}
\hfill
\begin{minipage}[t]{0.48\columnwidth}
\begin{tabular}{|c|c|c|}
\hline
$\alpha$ & ${\rm E_B},\,\, MeV$ & $\sqrt{<r^2>},\, \, fm$ \\
\hline       $-0.4$ & $-0.3 $         & $125 $   \\
\hline       $-0.6$ & $-8.3 $         & $26.7$   \\
\hline       $-0.8$ & $-28.4$         & $15.2$   \\
\hline
\end{tabular}
\caption{$|K \ \f \r$ ($|K \ \a \r$): the binding energies and
the size of the ground sate wave function for the potential in
Eq.~(\ref{V}) with $\kappa=140 \ MeV$.} \label{tabl2}
\end{minipage}
\end{table}

\begin{table}[ht]
\begin{minipage}[t]{0.48\columnwidth}
\begin{tabular}{|c|c|c|}
\hline
$\alpha$ & ${\rm E_B},\,\, MeV$ & $\sqrt{<r^2>},\, \, fm$ \\
\hline       $-0.2$ & $-0.6 $         & $41.1$   \\
\hline       $-0.4$ & $-25.5$         & $8.9 $   \\
\hline       $-0.6$ & $-90.0$         & $5.0 $   \\
\hline
\end{tabular}
\caption{$|D \ \D \r$: the binding energies and the size of the
ground sate wave function for the potential in Eq.~(\ref{V}) with
$\kappa=200 \ MeV$.} \label{tabl3}
\end{minipage}
\hfill
\begin{minipage}[t]{0.48\columnwidth}
\begin{tabular}{|c|c|c|c|}
\hline
$\alpha$ & ${\rm E_B},\,\, MeV$ & $\sqrt{<r^2>},\,\, fm$ \\
\hline       $-0.2$ & $-0.2 $     & $72.1$         \\
\hline       $-0.4$ & $-22.0$     & $8.8 $   \\
\hline       $-0.6$ & $-82.8$     & $5.3 $   \\
\hline
\end{tabular}
\caption{$|D \ \D \r$: the binding energies and the size of the
ground sate wave function for the potential in Eq.~(\ref{V}) with
$\kappa=220 \ MeV$.} \label{tabl4}
\end{minipage}
\end{table}

For this value of the coupling constant the binding energy of $|K\f \r$,
$|K\a\r$, $|\bar{K}\f \r$, $|\bar{K}\a\r$ systems ranges from about $8$ to
$20 \ MeV$ (for various values of $\kappa$), Tables~\ref{tabl1},
\ref{tabl2}. In the case of $|D\D \r$
the binding energy is about $80 - 90 \ MeV$,
Tables~\ref{tabl3}, \ref{tabl4}.

The potential given in Eq.~(\ref{V}) treats
$\f$ ($\a$), $K$, $D$ and $\D$ as though they were point-like
particles. The spatial
extent of the $K$ and $D$ mesons are approximately $0.4$ and $0.3
\ fm$. The $\f$($\a$) and $\D$ mesons are presumably even larger
(particularly if the interpretations (b) and (c) discussed in the Introduction
are correct).
Hence, the Yukawa potential in Eq.~(\ref{V}) cannot apply at
distances shorter than perhaps $0.5 \, fm$.  
The large width (short lifetimes), $\Gamma_{\f/\a}= 50 \, - \,
100 \ MeV$ and $\tau \sim (1.3 \, - \,0.7)\times 10^{-23} \ sec$,
makes observations of such states difficult. Roughly speaking,
since $\tau < T$, with $T$ being the time for completing one period in
the bound state, $T \sim 2 \, m_{\f}\,\ m_K /\kappa$, the $\f$ and
$\a$ decay before ``realizing'' that they are bound. The size of $\f$ ($\a$)
is of order of $1-2 \ fm$  and its velocity in traversing
the orbit is approximately $200\, MeV/500 \, MeV \approx 0.4$, so
that $T > (2.5\, - \,5) \times 10^{-23}$ seconds.

\begin{table}[ht]
\begin{minipage}[t]{0.48\columnwidth}
\begin{tabular}{|c|c|c|}
\hline
$\alpha$ & ${\rm E_B},\,\, MeV$ & $\sqrt{<r^2>},\, \, fm$ \\
\hline       $-0.4$ & $-8.71$         & $13.7$   \\
\hline       $-0.6$ & $-27.9$         & $8.5 $   \\
\hline       $-0.8$ & $-53.3$         & $6.4 $   \\
\hline
\end{tabular}
\caption{$|D \ \D \r$: the binding energies and the size of the
ground sate wave function for the potential in Eq.~(\ref{V}) for
$r>R$ and $V(r)=V(R)$ for $r<R$; $\kappa = 200 \ MeV,\ R=0.5 \
{\rm fm}$.} \label{tabl5}
\end{minipage}
\hfill
\begin{minipage}[t]{0.48\columnwidth}
\begin{tabular}{|c|c|c|c|}
\hline
$\alpha$ & ${\rm E_B},\,\, MeV$ & $\sqrt{<r^2>},\,\, fm$ \\
\hline       $-0.4$ & $-6.5 $         & $17.6$   \\
\hline       $-0.6$ & $-23.3$         & $8.9 $   \\
\hline       $-0.8$ & $-46.3$         & $7.0 $   \\
\hline
\end{tabular}
\caption{$|D \ \D \r$: the binding energies and the size of the
ground sate wave function for the potential in Eq.~(\ref{V}) for
$r>R$ and $V(r)=V(R)$ for $r<R$; $\kappa = 220 \ MeV,\ R=0.5 \
{\rm fm}$.} \label{tabl6}
\end{minipage}
\end{table}

\begin{table}[ht]
\begin{minipage}[t]{0.48\columnwidth}
\begin{tabular}{|c|c|c|}
\hline
$\alpha$ & ${\rm E_B},\,\,MeV$ & $\sqrt{<r^2>},\,\, fm$ \\
\hline       $-0.4$ & $-14.8$         & $10.5$   \\
\hline       $-0.6$ & $-46.9$         & $6.5 $   \\
\hline       $-0.8$ & $-91.0$         & $5.4 $   \\
\hline
\end{tabular}
\caption{$|D \ \D \r$: the binding energies and the size of the
ground sate wave function for the potential in Eq.~(\ref{V}) for
$r>R$ and $V(r)=V(R)$ for $r<R$; $\kappa = 200 \ MeV,\ R=0.3 \
{\rm fm}$.} \label{tabl7}
\end{minipage}
\hfill
\begin{minipage}[t]{0.48\columnwidth}
\begin{tabular}{|c|c|c|}
\hline
$\alpha$ & ${\rm E_B},\,\, MeV$ & $\sqrt{<r^2>},\,\, fm$ \\
\hline       $-0.4$ & $-4.9 $         & $17.5$   \\
\hline       $-0.6$ & $-16.7$         & $10.2$   \\
\hline       $-0.8$ & $-32.3$         & $7.9 $   \\
\hline
\end{tabular}
\caption{$|D \ \D \r$: the binding energies and the size of the
ground sate wave function for the potential in Eq.~(\ref{V}) for
$r>R$ and $V(r)=V(R)$ for $r<R$; $\kappa = 200 \ MeV,\ R=0.7 \
{\rm fm}$.} \label{tabl8}
\end{minipage}
\end{table}

In the case of the $|D\D \r$ ``molecule'' the spatial extent of
the stable $D$ and a very long lived $\D$ \cite{FN} can be
expected to reduce the binding energies. In this case the range
of the Yukawa potential, Eq.~(\ref{V}), is approximately $1 \ fm$.
The most conservative approach to the
unknown short range physics is to cut off the Yukawa potential at
distances shorter than, say, $R=0.5 \, fm$ and assume that
$V(r)=V(R)$ (the value of the potential in Eq.~(\ref{V}) at
$r=R$) for $r < R$. The binding energies and the
corresponding sizes of the wave functions are shown in
Tables~\ref{tabl5}, \ref{tabl6}, \ref{tabl7}, \ref{tabl8}, \ref{tabl9}, 
\ref{tabl10}.
As can be expected, there is a reduction in the
binding energies. The $|D \D \r$ system is still bound
by about $20 - 30 \ MeV$. The size of the bound state---of order of $8 \ fm$
is dominated by the tail of the $K$-exchange potential, Eq.~(\ref{V}).
We note in passing that in both cases the systems become unbound if
the potential is taken to be zero at $r < R$ (a radical
assumption).

\begin{table}[ht]
\begin{minipage}[t]{0.48\columnwidth}
\begin{tabular}{|c|c|c|}
\hline
$\alpha$ & ${\rm E_B},\,\, MeV$ & $\sqrt{<r^2>},\,\, fm$ \\
\hline       $-0.4$ & $-2.4 $         & $40.7$   \\
\hline       $-0.6$ & $-14.3$         & $20.4$   \\
\hline       $-0.8$ & $-35.2$         & $13.8$   \\
\hline
\end{tabular}
\caption{$|K \ \f \r$ ($|K \ \a \r$): the binding energies and
the size of the ground sate wave function for the potential in
Eq.~(\ref{V}) for $r>R$ and $V(r)=V(R)$ for $r<R$; $\kappa = 100 \
MeV,\ R=0.3 \ {\rm fm}$.} \label{tabl9}
\end{minipage}
\hfill
\begin{minipage}[t]{0.48\columnwidth}
\begin{tabular}{|c|c|c|}
\hline
$\alpha$ & ${\rm E_B},\,\, MeV$ & $\sqrt{<r^2>},\, \, fm$ \\
\hline       $-0.4$ &  $-1.9 $         & $48.3$   \\
\hline       $-0.6$ &  $-11.4$         & $20.6$   \\
\hline       $-0.8$ &  $-27.6$         & $14.2$   \\
\hline
\end{tabular}
\caption{$|K \ \f \r$ ($|K \ \a \r$): the binding energies and
the size of the ground sate wave function for the potential in
Eq.~(\ref{V}) for $r>R$ and $V(r)=V(R)$ for $r<R$; $\kappa = 100 \
MeV\,\ R=0.5 \ {\rm fm}$.} \label{tabl10}
\end{minipage}
\bigskip
\end{table}

Note that as the composite states overlap the strong short range
hyperfine interactions come into play since we have largely
different quarks in $\D$ and $D$ even  for the case (b) above
with $\D$ viewed as a four-quark construct. The tendency to form
these new loosely bound states would imply that at shorter
distances we have even stronger attraction that the extrapolation
of the relatively smooth Yukawa potential to short distances and
the results without any cutoff and a fortiori those in the case
(2) may be relevant!

It is interesting to note the drastic consequence of an even small attractive
scattering length---with no bound state in $K K$ (rather than $K \bar{K}$
channel). Arbitrary (sufficiently large) number of $K^0$ in a common
$S$-wave sate would then attract forming a condensate carrying macroscopic
strangeness {\it ala} Lee and Yang or Coleman's $Q$-balls \cite{SC}.
The longest range interaction between two kaons (and in fact any two mesons!)
due to the two pion exchange---specifically the $S$-wave projection thereof
in the $t$-channel is like a $\sigma$ ($J^{PC}=0^{++}$) or a scalar graviton
exchange
which is always attractive. The same also holds for $KN$ interactions.
However, the scattering
length in the Born approximation appropriate here is given by
$\int dr \ r^2 \ V(r)$ and the long range attraction is overcome (surely
for $K \ N$ from scattering data analysis and most likely for $KK$)
by the strong short range repulsion so that the condensates may not exist.

It is amusing to note in passing the (admittedly weak) connection
between the $|D \ \D \r$ bound state and the ``Efimov effect''
\cite{EF}. The latter (which inspired us to look at the present
problem) would arise for a zero energy $|KD \r$ $S$-wave bound
state and infinite scattering length. This in turn leads to an
infinite series of three body $|KDD \r$ bound states. The
ratios the binding energies and the sizes of the Efimov states
scale as $E_B(n+1)/ E_B(n) = e^{-2\pi} \sim 0.0016 \, , \,
<r>(n+1)/ <r>(n) = e^{\pi} \sim 25$ (see also \cite{RD,VK,BV}) .Clearly
in the present case where the range of the actual potential is
only about $1 \, fm$ this idealized case and the very extended---
$<r>(n) \sim 25^n$ (or contracted)---states in the above series
are irrelevant. Note, the the Yukawa potential, Eq.~(\ref{V})
goes to $1/r$ in the limit as $\e$ goes to infinity and $\kappa$
goes to zero rather rather then $1/r^2$ as in the Efimov effect.
The reason is that the Efimov effect requires exact
diagonalization of the degenerate perturbation transcending the
perturbative one-meson exchange \cite{VK,BV}.

\section{Experimental signatures}

Assuming that the $|D\D \r$ ``molecular'' state exists how can it be
produced and detected? Since the production requires two pairs of
$c\bar{c}$ quarks the
discussion in \cite{GN} is relevant, providing an upper bound on the
expected rate of the new loosely bound extended $|D\D \r$ state
which we term ${\cal M}$ for {\it molecular}.

The bound state should manifest as a narrow peak in the mass distribution of
associate $D$ and $\D$ decays. However, the limited experimental resolution
at BABAR and Fermi Lab experiments limits the extent that we can utilize
this. The different binding of $D^+$ and $D^0$ and different life-time
$\tau_{D^0} \sim \ 0.5 \ \tau_{D^{+}}$ may lead to some extra signatures in
specific charge dependence of the width and even in the binding energies.

\section{Acknowledgments}

T.C.  was supported in part by the U.S. Department of Energy under
Grant No. DE-FG02-93ER-40762.
B.G. was supported by the U.S. Department of Energy under
Grant No. DE-FG03-01ER-41196.
S.N. acknowledges a grant of the Israeli Academy of Science. 
B.G. and S.N. greatly acknowledge the hospitality of the Theory Group
for Quarks, Hadrons and Nuclei at the University of Maryland, College Park.


\begin{thebibliography}{99}
\bibitem{FN} The $\D \to D_s \pi$ decay mode is suppressed due to the isospin
non-conservation.
\end{thebibliography}
\end{document}